\documentclass[11pt,twoside]{article}

\usepackage{asp2006}
\usepackage{epsf}
\usepackage{psfig}
\usepackage{lscape}

\begin{document}
\title{Dust in External Galaxies}   
\author{Daniela Calzetti}   
\affil{Department of Astronomy, University of Massachusetts, Amherst, MA, USA}    

\begin{abstract} 
Existing (Spitzer Space Telescope) and upcoming (Herschel Space Telescope) facilities 
are deepening our understanding of the role of dust in tracing the energy budget and chemical evolution of galaxies. The tools we are developing while exploring the local Universe will in turn become pivotal 
in the interpretation of the high redshift Universe when near--future facilities (the Atacama Large Millimeter Array [ALMA], the Sub--Millimeter Array [SMA], the Large Millimeter Telescope [LMT], the James Webb Space Telescope [JWST]), and, possibly, farther--future ones, will begin operations.
\end{abstract}


\section{Introduction}  
Many of the questions left open today by the Cold Dark Matter (CDM) framework of galaxy formation and evolution are  nested into the basic understanding of the physical processes underlying star formation, as the baryonic matter sinks into the collapsing and merging dark matter haloes. These physical 
processes are   the `rules' for converting gas into stars and for driving the stellar and AGN feedback,  which affect the evolution of the luminous and non--luminous baryonic component of galaxies 
(this being, ultimately, what we observe). 

The open questions include, among others: the `missing satellites' problem, for which  CDM models predict about ten times, at least, more satellites around massive galaxies than what is actually 
observed \citep{Moore1999}; the `characteristic baryonic mass scale' of galaxies, 
for which the observed luminosity function of galaxies follows the Schechter functional shape and has a characteristic stellar mass of $\sim$6$\times$10$^{10}$~M$_{\sun}$ \citep{Kauffmann2003}, in stark 
contrast with the scale--free, power--law functional shape of the CDM dark halos mass function \citep[e.g., ][]{Benson2003}; the `angular momentum' (or `overcooling') problem, for which the sizes of CDM model 
galaxies are smaller than what observed \citep[e.g., ][]{Governato2006}; and the existence of bulge--less galaxies, again, difficult to produce within the CDM framework, where bulges are ubiquitous \citep{Mayer2008}.

Within this broad framework, the investigation of the infrared dust emission from galaxies offers a 
window on the questions revolving around star formation and feedback that is complementary to the 
UV--optical, `direct stellar light' approach. Indeed, about 50\% of the light observed in the Universe 
emerges at wavelengths longer than a few $\mu$m, and is due to dust--reprocessed stellar light 
\citep{Hauser2001,Dole2006}. More extreme still is the fraction of UV light, a probe of recent 
star formation, lost to dust absorption, this fraction being around 2/3--4/5 of the total UV from 
galaxies \citep{Calzetti2001}. 

\begin{figure}[!ht]
\plotone{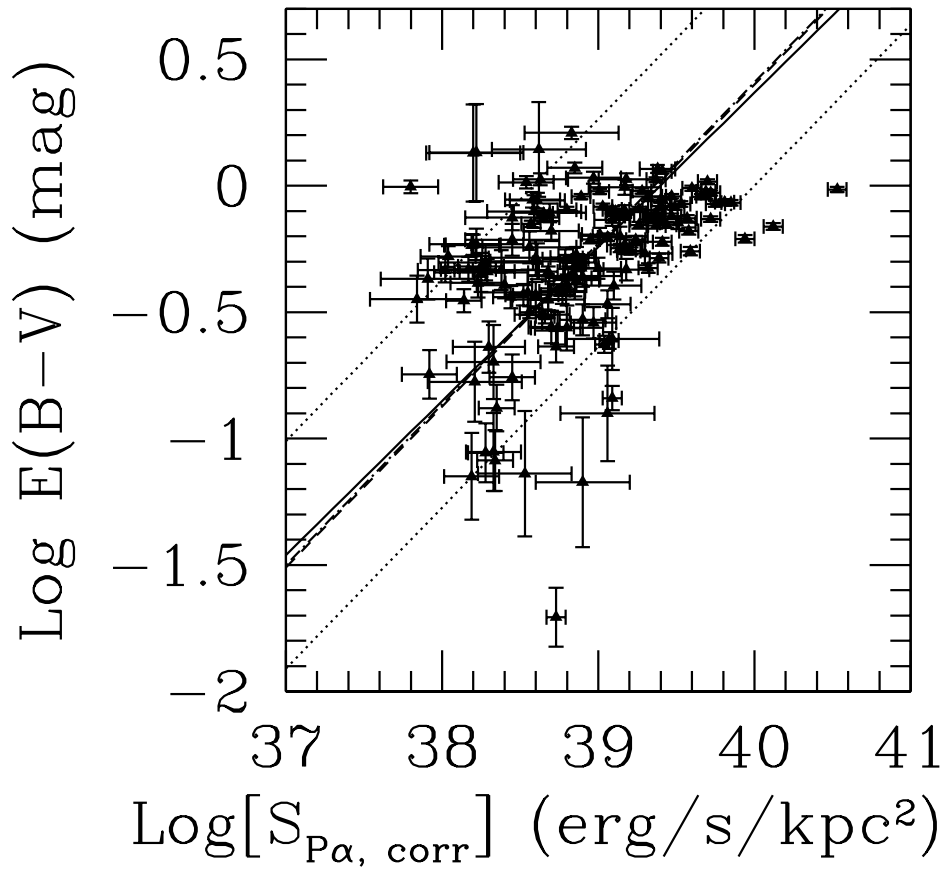}
\caption{The dust attenuation, expressed as the logarithm of the color excess E(B$-$V)=A$_V$/3.1 (in magnitudes),  as a function of the SFR density, in units of   M$_{\odot}$~yr$^{-1}$~kpc$^{-2}$, for 164 star--forming regions in 21 nearby galaxies. All the galaxies have metallicity close to 
the solar value. The continuous line is the expected trend after 
combining the Schmidt--Kennicutt law with the dust--to--gas ratio of the galaxies 
\citep{Calzetti2007}. The dotted lines are the 90\% boundary to the datapoints. \label{fig1}}
\end{figure}

The role of dust emission in tracing recent star formation becomes 
more important as the galaxy luminosity increases, since there is a loose correlation between
star formation rate (SFR) and the amount of dust extinction measured in galaxies and galaxy regions \citep[Figure~\ref{fig1}; e.g.,][]{Wang1996,Heckman1998,Calzetti2001,Hopkins2001,Sullivan2001,Buat2002,Calzetti2007}. 
The most luminous, and 
most strongly star forming, galaxies in the Universe do tend to be infrared--bright \citep[e.g.,][]{Sanders1996,Smail1997,Ivison1998,Blain2002}, even after taking into account AGN contributions 
\citep{Borys2005,Pope2008} and revisions to their infrared spectral energy distributions \citep[SEDs; ][]{Pope2006}. At the most basic level, accounting for the fraction of radiation from recent star formation 
absorbed by dust and re--processed in the infrared enables an accurate census of the total 
number of stars formed in galaxies across cosmic times, and a better understanding of the 
mass assembly of galaxies \citep[][and references therein]{Hopkins2006}. 

Furthermore, the investigation of the dust components present in the outflows of galaxies, such as 
the recently detected Polycyclic Aromatic Hydrocarbon (PAH) emission in the wind of the starburst galaxy 
M82 \citep{Engelbracht2006}, may 
help understand the energetics of stellar feedback, and shed light on the `pollutants' 
that can survive in the winds and possibly enter the intergalactic medium.  Stellar feedback has been  suggested as the `culprit' of the mass--metallicity relation in galaxies \citep{Garnett2002,Tremonti2004}, 
and the dust/gas ratio is correlated with the oxygen abundance in nearby star--forming galaxies 
\citep{Draine2007}. 

\section{Dust Emission and Dust Geometry}

One common application, when measuring star formation in 
external galaxies, is to use the far infrared (FIR) dust emission as a measure of the dust--absorbed 
UV light, which then enables recovering the `total' UV and quantifying the SFR as traced by the emission of young, massive stars. 
 
\begin{figure}[!ht]
\plotone{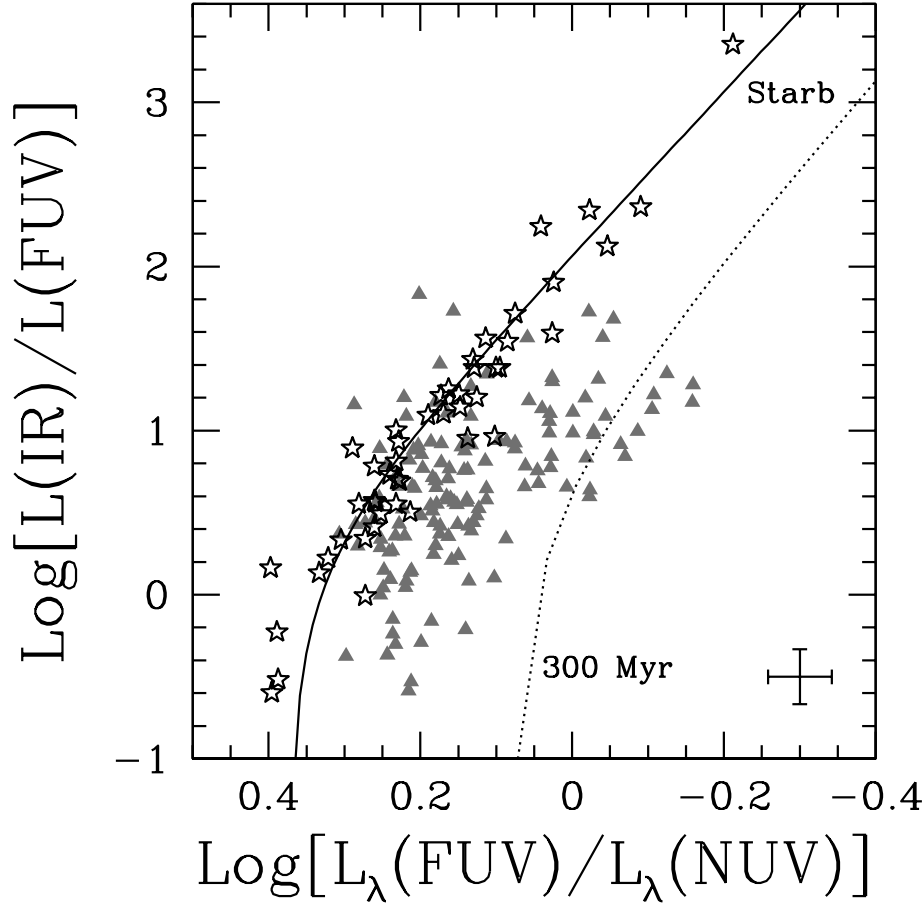}
\caption{The FIR/UV ratio (the ratio of the far--infrared to the far--UV luminosity) versus 
the UV~color (here expressed as the GALEX far--UV/near--UV 
color) for starburst galaxies and quiescently star--forming regions. The starburst galaxies are 
shown with star symbols. The grey filled triangles are star--forming regions within the 
galaxy NGC5194 \citep{Calzetti2005}. Redder UV colors correspond on average to larger FIR/UV 
ratios. The continuous line shows the best fit to the starburst galaxies, 
which is well represented by a model of a progressively more attenuated (from left to right) 
constant star--forming population. The dotted line shows the same dust attenuation trend 
for a 300~Myr old stellar population; this model represents a reasonable lower envelope 
to the NGC5194 star--forming regions, and to quiescent star--forming galaxies in general. \label{fig2}}
\end{figure}

Starburst galaxies show 
a well--defined correlation in the FIR/UV--versus--UV~color plane \citep[Figure~\ref{fig2}][]{Meurer1999}. 
The FIR/UV ratio is a measure of the UV attenuation suffered by the system: the higher the dust 
attenuation, the larger the amount of UV stellar energy reprocessed into the FIR by dust. Variations in 
the  UV~color probe the amount of dust reddening present in the system. Thus, in starbursts, 
 the UV reddening is a good tracer of the total UV dust attenuation \citep{Calzetti2001}. 
 
Quiescent star--forming galaxies and regions 
show a $\sim$10~times larger spread in the FIR/UV ratio than starburst galaxies, 
at fixed UV~color; the spread is in  the sense that the starburst galaxies form the upper 
envelope to the quiescently star--forming systems \citep[Figure~\ref{fig2} and][]{Buat2002,Buat2005,Bell2003,Gordon2004,Kong2004,Seibert2005,Calzetti2005}.  Thus, a far 
larger range of UV dust attenuations seems to be present in star--forming objects than in starbursts, 
for the same UV~color. This 
is possible if in quiescent star--forming systems the reddening of the UV~colors is not only a probe 
of dust, but also of ageing of the individual star--forming regions  contributing to the 
UV emission in the system \citep[regions up to an age of $\approx$100--300~Myr,][]{Calzetti2005}. 

An heuristic scenario can be built by recalling that, within starbursts, the large number of supernovae 
can cause the remnants to merge together and behave as a `collective entity', which can clear, 
thanks to the large input of mechanical energy, the interstellar medium within the starburst 
site and relocate the dust to the surrounding areas. In normal star--forming galaxies, 
such behavior is more fragmented and less energetic, as it takes place in individual HII regions, and the 
energy output is not as spatially concentrated as in starbursts. This can cause many of the young HII~regions to remain dust--enshrouded for longer times than in starbursts. Secular motions will 
eventually  separate them from their parent molecular cloud, but the stellar clusters will have aged 
 while migrating to areas of lower dust column density. The presence 
of multiple mechanisms for changing the relative geometry of dust and stars in normal star--forming galaxies, and the fragmented nature of their star formation (HII~regions behave like separate entities, 
and not like a collective one as in starbursts) is likely at the foundation of the observed spread 
in UV colors for constant  FIR/UV ratio. Aged clusters can still be bright enough (at least relative 
to their younger, but more dust enshrouded, counterparts) to contribute to the measured UV 
emission in the normal star--forming galaxies, yielding `red' UV colors (Figure~\ref{fig2}) not 
because of dust attenuation, but because of age. 

\section{The Infrared Emission for Tracing Star Formation}

The bolometric infrared emission from galaxies (integrated over the wavelength range 
$\lambda\sim$5--1000~$\mu$m) has been used as a 
SFR indicator at least since the data from the IRAS satellite showed that 
dust emission is widespread and significant in galaxies \citep{Soifer1986}.  
Young star--forming regions are dusty and the dust absorption 
cross--section peaks in the UV, i.e., in the same wavelength region where young, massive stars 
emission also peaks. Although the connection between SFR and infrared emission appears a 
straightforward one \citep[e.g.,][]{Devereux1992}, caveats have been raised for two major 
simplifications that underlie that connection \citep{Hunter1986,Persson1987,Rowan1989,Sauvage1992}. 
The first simplification is about the opacity of 
a galaxy: not all the luminous energy produced by recently formed stars is
re-processed by dust in the infrared; in this case, the FIR only recovers part of the SFR, and 
the fraction recovered depends, at least partially, on the amount of dust in the system, as well as 
on geometry. Recipes 
that account for the light of young, massive stars not absorbed by dust have been recently 
proposed \citep{Kennicutt2009}: these recipes combine the infrared luminosity with the observed 
H$\alpha$ luminosity, the latter a proxy for unobscured recent star formation, thus producing 
`hybrid' SFR indicators. 

The second simplification requires to neglect the heating of the dust by evolved, non--star forming 
populations; however, the latter will also contribute to the FIR emission, leading to an overestimate 
of the true SFR. If more evolved populations contribute mainly to the longer wavelength FIR, this 
extra contribution may be calibrated, at least for some classes of galaxies. A few approved observing projects for the upcoming Herschel Space Telescope plan to investigate this problem. 

The connection between infrared emission and SFR also provides the underlying physical mechanism for the correlation between FIR and radio emission in galaxies over 
5 orders of magnitude in luminosity \citep{Helou1985,Yun2001}. Recent analyses based on Spitzer 
data have shown that radio images are smoother versions of the infrared (70~$\mu$m) images 
of galaxies, with a correlation length that depends on the the age of the star formation event: cosmic 
rays created in more recent star formation events have not diffused significantly in the interstellar medium 
 of galaxies \citep{Murphy2006,Murphy2008}. The fundamental question of how two different 
processes, one related to the heating of dust by stars and the other affecting the propagation of the cosmic rays in galaxies, can produce a tight correlation over many orders of magnitude remains, however, open. The higher angular resolution images that will be provided by the Herschel 
Space Telescope in the FIR and by the EVLA in the radio will hopefully shed light on this 
fundamental question. 

\section{Mid--Infrared Emission from Galaxies}

Spitzer's high angular resolution (a few arcseconds) in the mid--infrared has offered the 
opportunity to investigate this wavelength range with unprecedented detail, thus building 
on the foundation laid by the Infrared Space Observatory (ISO). 

In particular, Spitzer has enabled a detailed investigation of the mid--infrared features in emission 
in the wavelength range $\sim$3--18~$\mu$m \citep[e.g.,][]{Smith2004,Engelbracht2006,Wu2006,Dale2006,Smith2007,Draine2007,Galliano2008b}. The dust emission in the 
mid--infrared ($\lambda\sim$5--40~$\mu$m)  range is characterized by both continuum and bands. 
The continuum is due to dust heated by a combination of single--photon and thermal equilibrium 
processes, with the latter becoming more and more prevalent over the former at longer wavelengths 
\citep[e.g.,][and references therein]{DraineLi2007}. The mid--infrared bands are generally attributed to Polycyclic Aromatic Hydrocarbons \citep[PAH, ][]{Leger1984,Sellgren1984}, large molecules 
transiently heated by single UV and optical photons in the general radiation field of galaxies or 
near B stars \citep{Li2002,Peeters2004,Matt2005}, and which can be
destroyed, fragmented, or ionized by harsh UV photon fields \citep{Boulanger1988,Pety2005}. 

Beyond being interesting in its own merit, the mid--infrared emission from galaxies has known 
renewed interest for its potential use as a SFR indicator in deep galaxy surveys, where 
the Spitzer (and Herschel) far--infrared bands correspond to rest--frame mid--infrared emission 
in high redshift galaxies \citep[e.g.,][]{Daddi2005,Marcillac2006,Daddi2007}. Furthermore, 
mono--chromatic SFR indicators avoid the uncertain extrapolations required by the bolometric 
infrared emission measurement, when only sparsely sampled dust SEDs are available (as often 
is the case for high--redshift data). The physical consideration underlying the use of the mid--infrared 
dust emission as a SFR indicator is  that the dust heated by hot, massive stars can have high 
temperatures and will preferentially emit at short infrared wavelengths. 

For the spectral regions where the dust SED is largely contributed by the PAH band emission, 
the definition of SFR indicators has been less than straightforward. The bands have shown to be heated 
by both recently formed \citep{Roussel2001,Forster2004,Dale2005,Dale2007} and evolved \citep{Haas2002,Boselli2004,Bendo2008} stellar populations. The PAH emission in 
galaxies shows a strong correlation with the emission of the cold dust heated by the general (non--star--forming) stellar population \citep{Haas2002,Bendo2008}. In particular, \citet{Bendo2008} finds that the 8~$\mu$m emission is more closely correlated with the 160~$\mu$m emission (cold dust) than the 24~$\mu$m emission (warm dust) on 2~kpc scales in galaxies. Analysis of the Spitzer 8~$\mu$m data of 
very nearby galaxies, like NGC300 and NGC4631, shows that the PAH band emission  highlights the rims of HII regions and is depressed inside the regions; this suggests that, even when closely associated with HII regions, the PAH dust is heated in the 
photo--dissociation regions  surrounding HII regions and likely destroyed within the HII~regions themselves  \citep{Helou2004,Bendo2006}. In general, the impact of the non--star--forming 
populations on the heating of the 8~$\mu$m emission is at the level of a factor of about 2, in the sense that the 8~$\mu$m Spitzer band will be affected roughly by this amount of uncertainty when used 
as a SFR indicator \citep{Calzetti2007}.

\begin{figure}[!ht]
\plotone{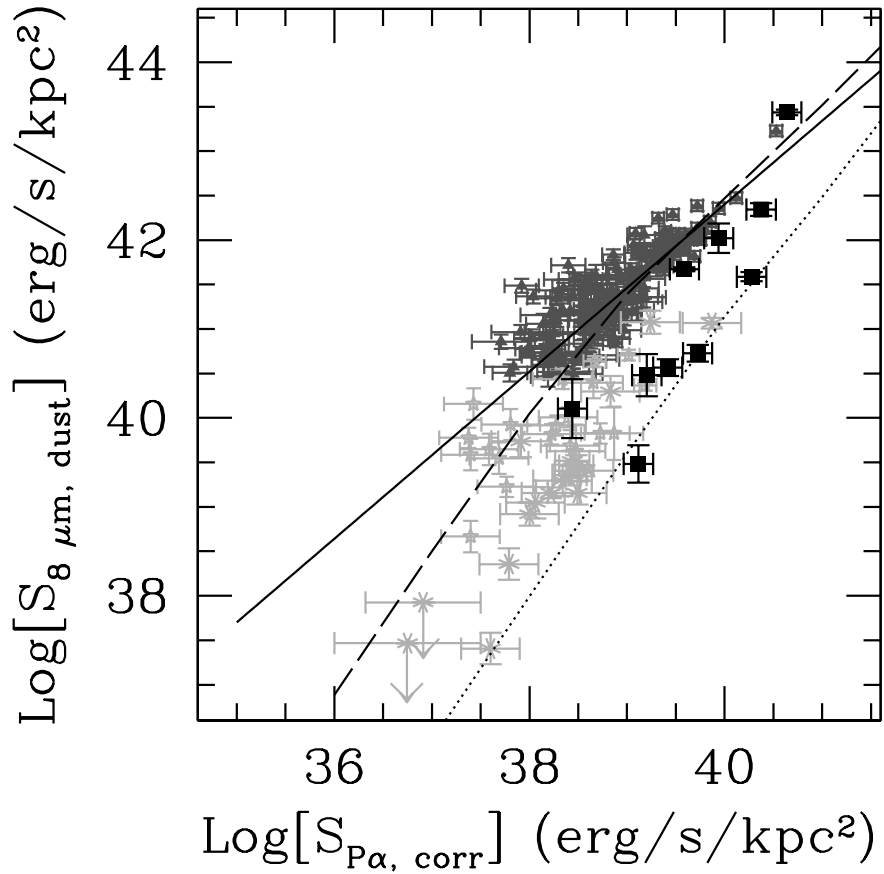}
\caption{The luminosity surface density (LSD=luminosity/area) at 8~$\mu$m versus the 
P$\alpha$ LSD of 10 starburst galaxies  and 220 star--forming regions in 33 nearby galaxies
\citep{Calzetti2007}. The 8~$\mu$m emission, from Spitzer images, has been corrected for stellar 
contribution. The P$\alpha$ line emission ($\lambda$=1.876~$\mu$m) 
has been corrected for dust extinction, and is used as an 
unbiased tracer of massive stars SFR. Of the 220 regions, the $\sim$180 regions in 
high--metallicity galaxies, 12$+$log(O/H)$>$8.3, are marked in dark grey, and the $\sim$40 regions in 
low--metallicity galaxies are  in light grey. The starburst galaxies \citep{Engelbracht2005} are low--metallicity ones (black squares). 
The continuous line is the best fit  to the high--metallicity star--forming regions (dark grey), with 
slope 0.94. Models for a young stellar population with increasing amount of star formation and dust 
are shown as a dash line (Z=Z$_{\odot}$) and a dot line (Z=1/10~Z$_{\odot}$), using the stellar 
population models of \citet{Leitherer1999} and the dust models of \citet{DraineLi2007}. 
The spread of the datapoints  
around the best fit line is well accounted for by a spread in the stellar population's age in the range 
2--8~Myr. \label{fig3}}
\end{figure}

Far more dramatic is the dependence of the PAH emission on metallicity \citep{Engelbracht2005,Rosenberg2006,Wu2006,Madden2006}. This dependence implies that regions with values about 1/10~Z$_{\sun}$ are about 10--30 times underluminous at 8~$\mu$m relative to regions 
with solar metallicity and same SFR \citep[Figure~\ref{fig3}, and][]{Calzetti2007,Rosenberg2008}. The observed correlation between the PAH `deficiency' and the hardness of the interstellar radiation field \citep{Madden2006} and the prevalence of 8~$\mu$m emission in the photo--dissociations regions surrounding the HII~regions 
\citep{Helou2004,Bendo2006} in nearby galaxies suggests that the strength/hardness of the  radiation 
 field influences the strength of the PAH emission, possibly by destroying or ionizing the carriers \citep{Madden2006,Wu2006,Galliano2008b}. A competing scenario suggests that in low--metallicity galaxies the deficiency  of PAH emission carriers may be due to delayed formation, rather than destruction \citep[e.g.,][]{Dwek2005,Galliano2008}.  In this scenario,  PAHs are formed in the envelopes around carbon--rich Asymptotic Giant Branch stars, later injected in the interstellar 
 medium over timescales of a few Gyr, while the other dust components are produced from supernova ejecta, which have timescales of a few to a few tens of Myr. If low--metallicity galaxies are 
 intrinsically young systems, there will be a delay between the formation of the PAH emission 
 carriers and the formation of all other dust components, thus accounting for the observed PAH emission `deficiency'. However, this scenario counters the results from Hubble Space Telescope observations that  
 show that even the supposedly least evolved galaxies in our local Universe harbor stellar populations that are at least a few Gyr old, and have been forming stars for at least that long \citep[e.g., Leo~A,][]{Cole2007}. Whatever the reason for the observed weakness of the PAH band emission in 
 low--metallicity systems, the impact on SFR indicators based on spectral regions where these bands 
 dominate the emission can be at the order--of--magnitude level, especially if applied to otherwise 
 unknown systems (as can be the case in high--redshift surveys). 

\begin{figure}[!ht]
\plotone{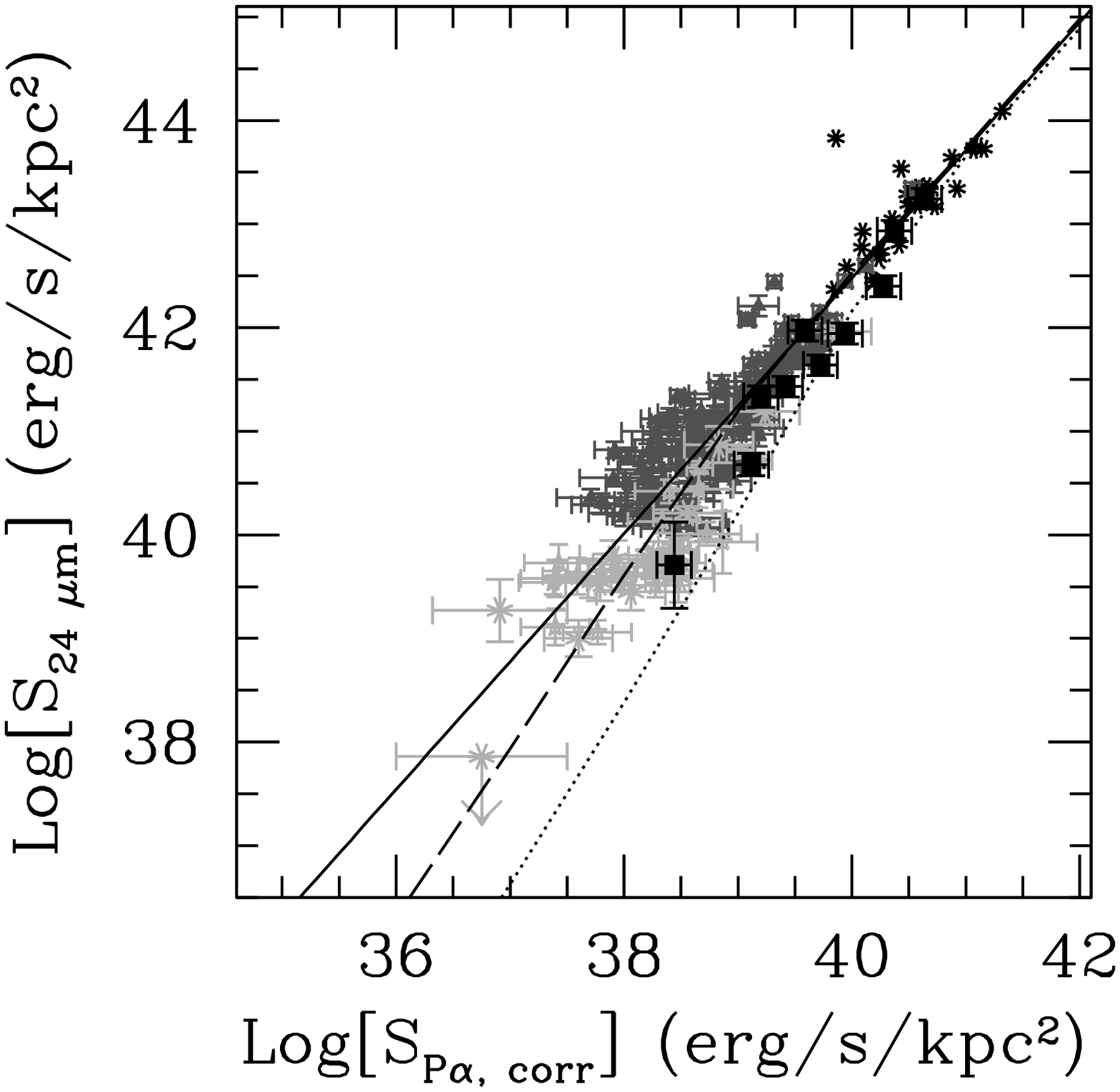}
\caption{The same as Figure~\ref{fig3}, with the vertical axis now reporting the 24~$\mu$m 
LSD, also from Spitzer. The black asterisks are the Luminous Infrared Galaxies (LIRGS) from \citet{AlonsoHerrero2006}. The best fit line (continuous line) has slope 1.2.\label{fig4}}
\end{figure}

Conversely, spectral regions located at longer wavelengths, where thermal equilibrium emission starts to dominate over single--photon processes, become increasingly more effective at tracing recent star formation. This is the case, for instance of the Spitzer 24~$\mu$m band, for which many calibrations exist in the literature, applicable to HII~regions, starburst--dominated galaxies, and normal 
star--forming galaxies \citep[Figure \ref{fig4}, and][]{Calzetti2005,Wu2005,AlonsoHerrero2006,PerezGonzalez2006,Calzetti2007,Rellano2007,Zhu2008,Rieke2009}. All of these calibrations assume that the objects which the SFR indicator is applied to are negligibly contaminated by AGN emission. The 24~$\mu$m band has a much lower sensitivity to metallicity than the 8~$\mu$m emission, 
decreasing by just a factor 2-4 for a tenfold decrease in metallicity. This small decrease is fully 
accounted for by the  increased transparency (lower dust--to--gas ratio) of the medium for lower 
metal abundances \citep{Calzetti2007}. 

\begin{figure}[!ht]
\plotone{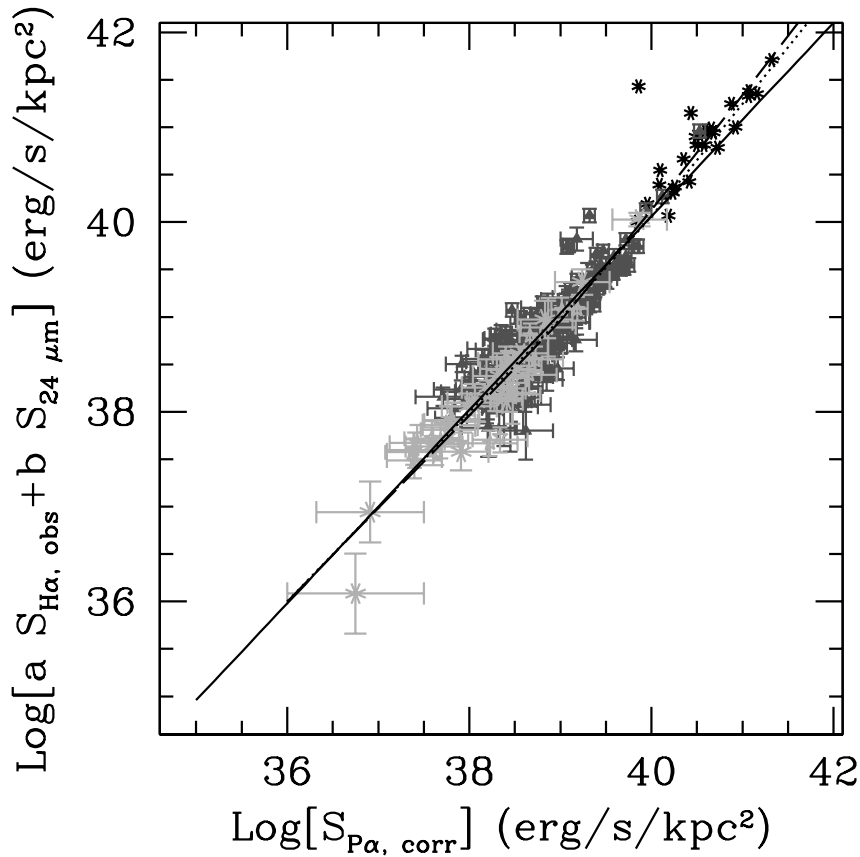}
\caption{As in Figure~\ref{fig3}, for the combined H$\alpha$ and 24~$\mu$m LSD. 
The data follow the 1--to--1 line up to the LIRGs, for b/a=0.031 \citep{Calzetti2007}. 
For this SFR indicator, data and models are degenerate in metallicity. Here the 24~$\mu$m emission 
traces the dust--obscured SFR, and the H$\alpha$ emission traces the unobscured SFR.  
\label{fig5}}
\end{figure}

As in the case of the FIR emission, SFRs based on the 24~$\mu$m emission rely on the assumption that most of the light from massive, young stars is re--processed by dust in the infrared. This is generally 
not the case in most galaxies, and `hybrid' SFR indicators are likely to be preferable to the use of a 
mono--chromatic SFR indicator. A hybrid SFR indicator that combines the observed H$\alpha$ recombination line emission and 24~$\mu$m dust emission can account for both 
the dust--obscured star formation (traced by the dust emission) and the unobscured portion of the star formation (traced by the ionizing photons) \citep[Figure~\ref{fig5}, and][]{Calzetti2007,Kennicutt2007,Zhu2008,Kennicutt2009}. The proportionality factor between the H$\alpha$ luminosity and the 24~$\mu$m emission changes from HII~regions/starburst--dominated galaxies to normal star--forming galaxies. This is probably a reflection of the fact that the underlying stellar population that provides most of the UV--optical photons (which heat the dust) changes from one type of system to the other \citep{Kennicutt2009}. 

\section{Summary}

ISO and Spitzer have enabled major progress in our understanding of the relation between 
the stellar populations that heat the dust and the various dust emission components present 
in the mid/far--infrared. 

The mid--infrared PAH emission is produced by dust grains transiently heated by both recently formed  and evolved stellar populations, thus the use of this emission to trace SFRs is subject to large (about a factor of 2 or so) uncertainties. The same band emission is sensitive to metallicity, showing variations of factors 10--30 for changes of a factor of 10 in metallicity. The application of SFR 
indicators based on those emission bands to distant objects should thus be performed with care. 

As we move to longer wavelengths, thermally heated dust increasingly contributes to the infrared emission. While still in the mid--infrared region ($\lambda<$40--50~$\mu$m), young, massive stars provide most of the heating for the dust, and this spectral region can be used reasonably well for tracing SFRs in galaxies. The main caveat here is that AGNs also contribute strongly to the heating of the dust emitting in the mid--infrared. Furthermore, all infrared--based SFR indicators need to rely on the assumption that most of the UV--optical light from stars is absorbed by dust, an assumption that is reasonably true only for luminous (L$>$10$^{10}$~L$_{\sun}$) galaxies.

The combination of Spitzer 70~$\mu$m and radio data has enabled the investigation of the FIR--radio correlation at higher angular resolution than any earlier data, leading to the understanding that radio maps are smoother versions of the infrared maps of galaxies, and the `diffusion' length of the cosmic rays correlates with the age of the star forming event. This is an area where the superior angular resolution of Herschel and EVLA will be able to unravel the underlying correlation between the processes that underlie the FIR and radio emission.

The Herschel Space Telescope, with its superior angular resolution in the far--infrared,  will also enable the investigation of the role of the different stellar populations in the heating of the different dust components that characterize the mid/far--infrared emission. These will, in turn, enable accurate determinations of SFR indicators using the long--wavelength infrared emission, of dust masses, 
local radiation field intensities, dust opacities and temperature distributions for nearby galaxies, and 
of the grain size distribution and composition.

The information on the characteristics of the dust emission and underlying dust properties that is being and will be, in the near future, obtained for nearby galaxies is going to provide the `toolbox' for the interpretation of distant galaxies that will be observed with Herschel, ALMA, SMA, and the USA--Mexico  50-meter millimeter telescope, the LMT. 


\acknowledgements 
This work would not have been possible without the many intellectual contributions of the SINGS 
(Spitzer Infrared Nearby Galaxies Survey) team. A list of team members can be found at:
http://sings.stsci.edu/Team/. 

This work is based in part on observations made with the Spitzer Space Telescope, which is operated by the Jet Propulsion Laboratory, California Institute of Technology under a contract with NASA.



\begin{thebibliography}{}
\bibitem[Alonso--Herrero et al.(2006)]{AlonsoHerrero2006} Alonso--Herrero, A., Rieke, G.H., Rieke, M.J., Colina, L., Perez-Gonzalez, P.G., \& Ryder, S.D. 2006, \apj, 650, 835
\bibitem[Bell(2003)]{Bell2003} Bell, E.F. 2003, \apj, 586, 794
\bibitem[Bendo et al.(2006)]{Bendo2006} Bendo, G.J., Dale, D.A., Draine, B.T., Engelbracht, C.W., 
Kennicutt, R.C., Calzetti, D., Gordon, K.D., Helou, G., Hollenbach, D., Li, AIgen, Murphy, E.J., et al. 
2006, \apj, 652, 283
\bibitem[Bendo et al.(2008)]{Bendo2008} Bendo, G.J., Draine, B.T., Engelbracht, C.W., Helou, G.,  Thornley, M.D., Bot, C., Buckalew, B.A., Calzetti, D., Dale, D.A., Hollenbach, D.J., et al. 2008 \mnras, 389, 629
\bibitem[Benson et al.(2003)]{Benson2003} Benson, A.J., Bower, R.G, Frenk, C.S., Lacey, C.G., Baugh, C.M., \& Cole, S. 2003, \apj, 599, 38
\bibitem[Blain et al.(2002)]{Blain2002} Blain, A.W., Smail, I., Ivison, R.J., Kneib, J.-P., \& Frayer, D.T. 
2002, Ph Rev. 369, 111
\bibitem[Borys et al.(2995)]{Borys2005} Borys, C., Smail, I., Chapman, S.C., Blain, A.W., Alexander, 
D.M., \& Ivison, R.J. 2005, \apj, 635, 853
\bibitem[Boselli, Lequeux \& Gavazzi(2004)]{Boselli2004} Boselli, A., Lequeux, J.,
  \& Gavazzi, G. 2004, \aap, 428, 409
 \bibitem[Boulanger et al.(1988)]{Boulanger1988} Boulanger, F., Beichmann, C.,
Desert, F.--X., Helou, G., Perault, M., \& Ryter, C. 1988, \apj, 332, 328
\bibitem[Buat et al.(2002)]{Buat2002} Buat, V., Boselli, A., Gavazzi, G., \&  Bonfanti, C. 2002, \aap, 383, 801
\bibitem[Buat et al.(2005)]{Buat2005} Buat, V., Iglesias-Paramo, J., Seiber, M., Burgarella, D., 
Charlot, S., Martin, D.C., Xu, C.K., Heckman, T.M.,, Boissier, S., Boselli, A., et al.  2005, \apj, 619, L51 
\bibitem[Calzetti(2001)]{Calzetti2001} Calzetti, D. 2001, \pasp, 113, 1449
\bibitem[Calzetti et al.(2000)]{Calzetti2000} Calzetti, D., Armus, L., Bohlin, R.C., Kinney, A.L., Koornneef, J., \& Storchi-Bergmann, T. 2000, \apj, 533, 682
\bibitem[Calzetti et al.(2005)]{Calzetti2005} Calzetti, D., Kennicutt, R.C., Bianchi, L., Thilker, D.A., Dale, D.A., Engelbracht, C.W., Leitherer, C.,  Meyer, M.J., et al. 2005, \apj, 633, 871
\bibitem[Calzetti et al.(2007)]{Calzetti2007} Calzetti, D., Kennicutt, R.C., 
Engelbracht, C.W., Leitherer, C., Draine, B.T., Kewley, L., Moustakas, J., Sosey, M., Dale, D.A., 
Gordon, K.D.,  et al. 2007, \apj, 666, 870
\bibitem[Cole et al.(2007)]{Cole2007}  Cole, A.A., Skillman, E.D., Tolstoy, E. Gallagher, J.S., 
Aparicio, A., Dolphin, A.E., Gallart, C., Hidalgo, S.L., Saha, A., Stetson, P.B., \& Weisz, D.R. 2007, 
\apj, 659, L17
\bibitem[Daddi et al.(2005)]{Daddi2005} Daddi, E., Dickinson, M., Chary, R., Pope, A., Morrison, G., Alexander, D.M., Bauer, F.E., Brandt, W.,N., Giavalisco, M., Ferguson, H., et al. 2005, \apj, 631, L13
\bibitem[Daddi et al.(2007)]{Daddi2007} Daddi, E., Dickinson, M., Morrison, G., Chary, R., Cimatti, A., Elbaz, D., Frayer, D. Renzini, A., Pope, A., Alexander, D.M., et al. 2007, \apj, 670, 156
\bibitem[Dale et al.(2005)]{Dale2005} Dale, D.A., Bendo, G.J., Engelbracht, C.W., Gordon, K.D., Regan, M.W., Armus, L., Cannon, J.M., Calzetti, D., Draine, B.T., Helou, G., et al. 2005, \apj, 633, 857
\bibitem[Dale et al.(2007)]{Dale2007} Dale, D.A., Gil de Paz, A., Gordon, K.D., Hanson, H.M., Armus, L.,  Bendo, G.J., Bianchi, L., Block, M., Boissier, S., Boselli, A., et al. 2007, \apj, 655, 863
\bibitem[Dale et al.(2006)]{Dale2006} Dale, D.A., Smith, J.D.T., Armus, L., Buckalew, B.A., Helou, G., Kennicutt, R.C., Moustakas, J., Roussel, H., Sheth, K., Bendo, G.J., et al. 2006, \apj, 646, 161
\bibitem[Devereux \& Young(1992)]{Devereux1992} Devereux, N.A., \& Young, J.S. 1992, \aj, 103, 1536
\bibitem[Dole et al.(2006)]{Dole2006} Dole, H., Lagache, G., Puget, J.-L., Caputi, K.I., Fern‡ndez-Conde, N., Le Floc'h, E., Papovich, C., PŽrez-Gonz‡lez, P.G., Rieke, G.H., \& Blaylock, M. 2006, \aap, 451, 417
\bibitem[Draine \& Li(2007)]{DraineLi2007} Draine, B.T., \& Li A. 2007, ApJ, 657, 810. 
\bibitem[Draine et al.(2007)]{Draine2007} Draine, B.T., Dale, D.A., Bendo, G., Gordon, K.D., Smith, 
J.D.T., Armus, L., Engelbracht, C.W., Helou, G., Kennicutt, R.C., Li, A., et al.  2007, \apj, 633, 866. 
\bibitem[Dwek(2005)]{Dwek2005} Dwek, E., 2005, AIPC, 804, 197
\bibitem[Engelbracht et al.(2005)]{Engelbracht2005} Engelbracht, C.W., Gordon, K.D., Rieke, G.H., Werner, M.W., Dale, D.A., \& Latter, W.B.  2005, ApJ, 628, 29
\bibitem[Engelbracht et al.(2006)]{Engelbracht2006} Engelbracht, C.W., Kundurthy, P., Gordon, K.D., Rieke, G.H., Kennicutt, R.C., Smith, J.-D. T., Regan, M.W., Makovoz, D., Sosey, M., Draine, B.T., 
et al. 2006, ApJ, 642, L127
\bibitem[F\"orster Schreiber et al.(2004)]{Forster2004} F\"orster Schreiber, N.M.,  Roussel, H., Sauvage, M., \& Charmandaris, V., 2004, \aap, 419, 501
\bibitem[Galliano et al.(2008a)]{Galliano2008} Galliano, F., Dwek, E., \& Chanial, P. 2008, \apj, 672, 214
\bibitem[Galliano et al.(2008b)]{Galliano2008b} Galliano, F., Madden, S.C., Tielens, A.G.G.M., 
Peeters, E., \& Jones, A.P. 2008, \apj, 679, 310
\bibitem[Garnett(2002)]{Garnett2002} Garnett, D.R. 2002, \apj, 581, 1019
\bibitem[Gordon et al.(2004)]{Gordon2004} Gordon, K.D., Perez--Gonzalez,
P.G., Misselt, K.A., Murphy, E.J., Bendo, G.J., Walfer, F., Thornley, M.D., Kennicutt, R.C., et al. 2004, \apjs 154, 215 
\bibitem[Governato(2006)]{Governato2006} Governato, F. 2006, in The Fabulous Destiny of Galaxies: 
Bridging Past and Present, Proceedings of the Vth Marseille International Cosmology 
COnference eds. V. LeBrun, A. Mazure, S. Arnouts amd D. Burgarella (Paris: Frontier Group), 241
\bibitem[Haas, Klaas \& Bianchi(2002)]{Haas2002} Haas, M., Klaas, U., \& Bianchi, S. 2002, \aap, 385, L23
\bibitem[Hauser \& Dwek(2001)]{Hauser2001} Hauser, M.G., \& Dwek, E. 2001, \araa, 39, 249
\bibitem[Heckman et al.(1998)]{Heckman1998} Heckman, T.M., Robert, C., Leitherer,
  C., Garnett, D.R., \& van der Rydt, F. 1998, \apj, 503, 646
\bibitem[Helou, Soifer, \& Rowan--Robinson(1985)]{Helou1985} Helou, G.X., Soifer, B.T., \& 
Rowan--Robinson, M. 1985, \apj, 298, L7
\bibitem[Helou et al.(2004)]{Helou2004} Helou, G., Roussel, H., Appleton, P.,
Frayer, D., Stolovy, S., Storrie--Lombardi, L., Hurst, R., Lowrance, P., et al. 2004, \apjs, 154, 253
\bibitem[Hopkins \& Beacom(2006)]{Hopkins2006} Hopkins, A.M., \& Beacom, J.F. 2006, \apj, 651, 142
\bibitem[Hopkins et al.(2001)]{Hopkins2001} Hopkins, A.M., Connolly, A.J., Haarsma, D.B., \& 
Cram, L.E. 2001, \aj, 122, 288
\bibitem[Hunter et al.(1986)]{Hunter1986} Hunter, D.A., Gillett, F.C., Gallagher, J.S., Rice, W.L., \& Low, 
F.J. 1986, \apj, 303, 171
\bibitem[Ivison et al.(1998)]{Ivison1998} Ivison, R.J., Smail, I., Le Borgne, J.-F., Blain, A.W., Kneib, J.-P., Bezecourt, J., Kerr, T.H., \&  Davies, J.K. 1998, \mnras, 298, 583
\bibitem[Kauffmann et al.(2003)]{Kauffmann2003} Kauffmann, G., Heckman, T.M., White, S.D.M., Charlot, 
S., Tremonti, C., et al. 2003, MNRAS, 341, 33
\bibitem[Kennicutt et al.(2007)]{Kennicutt2007} Kennicutt, R.C., Calzetti, D., Walter, F., Helou, G.,m Hollenbach, D., Armus, L., Bendo, G., Dale, D.A., Draine, B.T., Engelbracht, C.W., et al. 2007a, \apj, 671, 333
\bibitem[Kennicutt et al.(2009)]{Kennicutt2009} Kennicutt, R.C., Hao, C., Calzetti, D., et al. 2009, in prep. 
\bibitem[Kong et al.(2004)]{Kong2004} Kong, X., Charlot, S., Brinchmann,
J., \& Fall, S.M. 2004, \mnras, 349, 769
\bibitem[Leger \& Puget(1984)]{Leger1984} Leger, A., \& Puget, J.L. 1984,
\aap, 137, L5
\bibitem[Leitherer et al.(1999)]{Leitherer1999} Leitherer, C., Schaerer, D., 
Goldader, J.D., Gonz\'alez Delgado, R.M., Robert, C., Kune, D.F., de Mello, 
D.F., Devost, D., \& Heckman, T.M. 1999, \apjs, 123, 3
\bibitem[Li \& Draine(2002)]{Li2002} Li, A., \& Draine, B.T. 2002, \apj, 572, 762
\bibitem[Lonsdale Persson \& Helou(1987)]{Persson1987} Lonsdale Persson, C.J., \& Helou, G.X. 
1987, \apj, 314, 513
\bibitem[Madden et al.(2006)]{Madden2006} Madden, S.C., Galliano, F., Jones, A.P.,\& Sauvage, M. 
2006, \aap, 446, 877
\bibitem[Marcillac et al.(2006)]{Marcillac2006} Marcillac, D., Elbaz, D., Chary, R.R., Dickinson, M.,  Galliano, F., \& Morrison, G. 2006, \aap, 451, 57
\bibitem[Mattioda et al. (2005)]{Matt2005} Mattioda, A.L., Allamandola, L.J., \& Hudgins, D.M. 
 2005, \apj, 629, 1183
 \bibitem[Mayer et al.(2008)]{Mayer2008} Mayer, L., Governato, F., \& Kaufmann, T. 2008, to 
 appear in Advanced Science Letters (astroph/0801.3845)
\bibitem[Meurer, Heckman \& Calzetti(1999)]{Meurer1999} Meurer, G.R., Heckman,
  T.M., \& Calzetti, D. 1999, \apj, 521, 64 
\bibitem[Moore et al.(1999)]{Moore1999} Moore, B., Ghigna, S., Governato, F.,  Lake, G., Quinn, T.,
  Stadel, J., \& Tozzi, P. 1999, \apj, 524, L19
\bibitem[Murphy et al.(2006)]{Murphy2006} Murphy, E.J., Helou, G., Braun, R., Kenney, J.D.P., Armus, L., Calzetti, D., Draine, B.T., Kennicutt, R.C., Roussel, H., Walter, F., et al. 2006, \apj, 651, L111
\bibitem[Murphy et al.(2008)]{Murphy2008} Murphy, E.J., Helou, G., Kenney, J.D.P., Armus, L., \& Braun, R. 2008, \apj, 678, 828
\bibitem[Peeters, Spoon \& Tielens(2004)]{Peeters2004} Peeters, E., Spoon,
H.W.W., \& Tielens, A.G.G.M. 2004, \apj, 613, 986
\bibitem[Perez--Gonzalez et al.(2006)]{PerezGonzalez2006} Perez--Gonzalez, P.G., Kennicutt, R.C., Gordon, K.D., Misselt, K.A., Gil de Paz, A., Engelbracht, C.W., Rieke, G.H., Bendo, G.J., Bianchi, L., Boissier, S., Calzetti, D., Dale, D.A., et al.  2006, \apj, 648, 987
\bibitem[Pety et al.(2005)]{Pety2005} Pety, J., Teyssier, D., Fosse`,
D., Gerin, M., Roueff, E., Abergel, A., Habart, E., \& Cernicharo, J. 2005, \aap, 435, 885
\bibitem[Pope et al.(2008)]{Pope2008} Pope, A., Chary, R.-R., Alexander, D.M., Armus, L., Dickinson, M., Elbaz, D., Frayer, D., Scott, D., \& Teplitz, H. 2008, \apj, 675, 1171
\bibitem[Pope et al.(2006)]{Pope2006} Pope, A., Scott, D., Dickinson, M., Chary, R.-R., Morrison, G.,  Borys, C., Sajina, A., Alexander, D.M., Daddi, E., Frayer, D., MacDonald, E., \& Stern, D. 2006, \mnras, 
370, 1185
\bibitem[Rela\~no et al.(2007)]{Rellano2007} Rela\~no, M., Lisenfeld, U., Perez-Gonzalez, P.G., 
  Vilchez, J.M., \& Battaner, E. 2007, \apj, 667, L141
\bibitem[Rieke et al.(2009)]{Rieke2009} Rieke, G.H., Alonso-Herrero, A., Weiner, B.J., Perez--Gonzalez, P.G., Blaylock, M., Donley, J.L, \& Marcillac, D. 2009, \apj, in press (astroph/0810.4150)
\bibitem[Rosenberg et al.(2006)]{Rosenberg2006} Rosenberg, J.L., Ashby, M.L.N., Salzer, J.J., Huang, J.-S. 2006, \apj, 636, 742
\bibitem[Rosenberg et al.(2008)]{Rosenberg2008} Rosenberg, J.L., Wu, Y., Le Floc'h, E., Charmandaris, V. Ashby, M.L.N., Houck, J.R., Salzer, J.J., Willner, S.P. 2008, \apj, 674, 814
\bibitem[Roussel et al.(2001)]{Roussel2001} Roussel, H., Sauvage, M., Vigroux, L., \& Bosma, A. 2001, \aap, 372, 427
\bibitem[Rowan--Robinson \& Crawford(1989)]{Rowan1989} Rowan--Robinson, M., \& Crawford, 
J. 1989, \mnras, 238, 523
\bibitem[Sanders \& Mirabel(1996)]{Sanders1996} Sanders, D.B., \& Mirabel, I.F. 1996, \araa, 34, 749
\bibitem[Sauvage \& Thuan(1992)]{Sauvage1992} Sauvage, M., \& Thuan, T.X. 1992, \apj, 396, L69
\bibitem[Seibert et al.(2005)]{Seibert2005} Seibert, M., Martin, D.C., Heckman, T.M., Buat, V., Hoopes, C.,  Barlow, T., Bianchi, L., Byun, Y.-I., Donas, J., Forster, K., et al. 2005, \apj, 619, L23
\bibitem[Sellgren(1984)]{Sellgren1984} Sellgren, K. 1984, \apj, 277, 623
\bibitem[Smail, Ivison, \& Blain(1997)]{Smail1997} Smail, I., Ivison, R.J., \& Blain, A.W. 1997, ApJ, 490, L5
\bibitem[Smith et al.(2004)]{Smith2004} Smith, J.D.T., Dale, D.A., Armus, L., Draine, B.T., Hollenbach, D.J., Roussel, H., Helou, G., Kennicutt, R.C., Li, A., Bendo, G.J., et al. 2004, \apjs, 154, 199
\bibitem[Smith et al.(2007)]{Smith2007}  Smith, J.D.T., Draine, B.T., Dale, D.A., Moustakas, J., Kennicutt, R.C., Helou, G., Armus, L., Roussel, H., Sheth, K., Bendo, G.J., et al. 2007, \apj, 656, 770
\bibitem[Soifer et al.(1986)]{Soifer1986} Soifer, B.T., Sanders, D.B., Neugebauer, G., Danielson, G.E., Lonsdale, C.J., Madore, B.F., \&  Persson, S.E. 1986, \apj, 303, L41
\bibitem[Sullivan et al.(2001)]{Sullivan2001} Sullivan, M., Mobasher, B., Chan, B., Cram, L., Ellis, R., 
Treyer, M., \& Hopkins, A. 2001, \apj, 558, 72
\bibitem[Tremonti et al.(2004)]{Tremonti2004} Tremonti, C.A., Heckman, T.M., Kauffmann, G.,  Brinchmann, J., Charlot, S., White, S.D.M., Seibert, M., Peng, E.W., Schlegel, D.J., Uomoto, A., et al. 2004, \apj, 613, 898
\bibitem[Wang \& Heckman(1996)]{Wang1996} Wang, B., \& Heckman, T.M. 1996, \apj,
  457, 645
\bibitem[Wu et al.(2005)]{Wu2005}  Wu, H., Cao, C., Hao, C.-N., Liu, F.-S., Wang, J.-L., Xia, X.-Y., Deng, Z.-G., \& Young, C. K.-S. 2005, \apj, 632, L79
\bibitem[Wu et al.(2006)]{Wu2006} Wu, Y., Charmandaris, V., Hao, L., Brandl, B.R., Bernard-Salas, J., Spoon, H.W.W., \& Houck, J.R. 2006, \apj, 639, 157
\bibitem[Yun, Reddy, \& Condon(2001)]{Yun2001} Yun, M.S., Reddy, N.A., \& Condon, J.J. 2001, \apj, 554, 803
\bibitem[Zhu et al.(2008)]{Zhu2008} Zhu, Y.-N., Wu, H., Cao, C., \& Li, H.-N. 2008, \apj, 686, 155
\end{thebibliography}
\end{document}